# OpenRIS: Democratizing reconfigurable intelligent surfaces for real-world wireless enhancements

Weicong Chen[1], Junjie Ai[1], Lin Bai[2], Xiaokun Teng[1], Wen Jun Teng[1], Wankai Tang[1], Xiao Li[1*], Wei Xiang Jiang[2], Shi Jin[1]*, and Tie Jun Cui[2]*

[1]National Mobile Communications Research Laboratory, Southeast University, Nanjing 210096, China

[2]State Key Laboratory of Millimeter Waves, Southeast University, Nanjing 210096, China

*E-mail: li_xiao@seu.edu.cn; jinshi@seu.edu.cn; tjcui@seu.edu.cn

## Abstract

Wireless enhancement is critical for next-generation mobile communication systems to realize seamless connectivity, yet traditional network expansion strategies are becoming economically unsustainable. Reconfigurable intelligent surfaces (RISs) provide a promising alternative by improving signal utilization. However, high hardware and deployment costs of advanced RISs limit their large-scale application. Here, we democratize this technology with OpenRIS, an open-source and low-cost platform composed of Lego-like meta-bricks. With digital-twin assistance, these meta-bricks can be flexibly assembled into arbitrary shapes to achieve customized, mass-deployable wireless enhancement without extra power. Experiments and full-wave simulations verify that the discretized OpenRIS achieves consistent performance with the continuous RIS. We further develop a dual-user wireless transmission system and a three-dimensional coverage measurement system to showcase the versatile applicability of OpenRIS in wireless enhancements. As a plug-and-play solution, OpenRIS accelerates the translation of RIS theory into practice and is poised to integrate into infrastructure, reshaping the future wireless world as steel and concrete shape modern cities.

The evolution of mobile communication systems has greatly driven the progress of human civilization, establishing a critical digital infrastructure that underpins personal communications[1], agricultural monitoring[2], industrial automation[3], and military command[4], and so on. As emerging applications such as integrated sensing and communications[5], telemedicine[6], and virtual/augmented reality[7] continue to mature, wireless communications are becoming an indispensable cornerstone of future intelligent society. A typical wireless communication system comprises three fundamental components: transmitter, wireless channel, and receiver. Since the inception of wireless communication, technological advancements have predominantly focused on optimizing transceivers[8]-[12], while the uncontrollable wireless channel[13][14] has remained a fundamental bottleneck. The lack of active control over the wireless propagation medium severely restricts the performance of wireless transmissions[15]. Conventional transceiver designs for wireless enhancements rely heavily on resource-intensive expansion to achieve incremental performance gains. These brute-force scale-up approaches lead to high cost, excessive energy consumption, and low spectral efficiency, rendering traditional network expansion strategies increasingly economically unsustainable.

To address this long-standing issue, reconfigurable intelligent surfaces (RISs)[16]-[19] have emerged as a disruptive wireless enhancement technology that repurposes existing signals. Derived from metasurface technology[20]-[21], RIS enables dynamic control of electromagnetic wave propagation, effectively reshaping wireless environments[22]-[26] for more efficient signal utilization. Originally proposed as a low-cost, low-power wireless enhancement solution, RIS now faces practical deployment hurdles driven by diverse functional demands. The pursuit of high-precision reflection/refraction phase control[26], non-diagonal configurations[27], intelligent adaptation[28], and integrated sensing[29] and amplification[30] capabilities has inadvertently escalated its hardware complexity, cost, and power consumption. While advanced RISs offer unparalleled flexibility in wireless signal manipulation, their over-engineered capabilities often prove excessive for the most common commercial applications requiring sustained coverage enhancement, undermining their original promise. Moreover, compared to extensive academic exploration[31]-[36], practical RIS applications[37]-[40] remain scarce due to prohibitive manufacturing costs and intricate fabrication processes. The gap between theoretical research and real-world implementation further impedes RIS commercial large-scale deployment.

To realign RIS with its foundational goals—simplicity, affordability, and energy efficiency—we present OpenRIS, an open-source, ultra-low-cost, and zero-energy platform. By discretizing the RIS into Lego-like meta-bricks, OpenRIS achieves deploy-and-forget electromagnetic reconfigurability without relying on continuous power supply to sustain electronic reconfiguration, in sharp contrast to conventional RIS. This design distinguishes OpenRIS from passive fixed metasurfaces[41]-[44] (limited by static electromagnetic configuration) and active programmable counterparts[37]-[40] (often constrained by high deployment costs and physical dimensions), through its innovative plug-and-play and replaceable modular architecture. By leveraging digital twins to capture the characteristics of the target wireless environment and constituent meta-bricks, we can rapidly prototype large-scale OpenRIS arrays of arbitrary shape on-site for predefined wireless coverage enhancement—while also validating their performance and adapting array dimensions as coverage requirements evolve. This design philosophy democratizes RIS technology with significantly reduced manufacturing and deployment costs, bridging the gap between theoretical analysis and practical implementation for wireless enhancement.

The urgency for such innovations is underscored by persistent coverage gaps in existing commercial wireless networks, particularly in millimeter-wave (mmWave) bands where deploying additional base stations for the "last-mile" wireless coverage enhancement incurs prohibitive costs. By dynamically tuning the wireless channel for more efficient signal propagation, OpenRIS offers a scalable alternative. In this work, we report a mmWave OpenRIS design and develop a dual-user wireless transmission system and a three-dimensional coverage measurement system to showcase the versatile applicability of OpenRIS in wireless enhancements. Experiments demonstrate that a conformal OpenRIS establishes stable links for out-of-coverage users, while its reconfigured planar form enables uniform 3D coverage—both achieving an average signal gain of ~ 20 dB at only 1% of the cost of conventional RIS solutions. The flexibility of OpenRIS positions it as an easily accessible wireless enhancement platform for future wireless systems, where reconfigurable wireless channels will be critical to balancing wireless transmission performance, cost, and sustainability.

## Results

### OpenRIS enabled wireless enhancement

Figure 1 schematically illustrates two representative use cases of OpenRIS on wireless enhancement: steering beams for directional uplink hotspot enhancement and dispersing narrow beam for downlink 3D wide coverage. Millimeter-wave communications face inherent challenges in uplink due to the limited transmit power of user equipment (UE), the severe path loss, and narrow angular coverage of the access point (AP). When hotpot UEs like high-definition (HD) cameras or extended reality (XR) equipment are temporarily installed at fixed positions with large angular offsets relative to the AP, they cannot establish high-speed transmission link for uploading data through either direct paths or environmental reflection paths. RISs can provide artificial non-specular reflection path via manipulating the wave-front of EM wave, however, conventional planar RIS cannot cater to walls with arbitrary shape. The proposed meta-bricks can constitute any conformal OpenRIS like Fig. 1(a) to redirect signals from UEs outside of coverage to the AP, thereby enhancing directional hotspot uplink coverage. In 3D communication scenarios such as stairwells, wireless transmission requirement may shift from ultra-high-speed data transmission to maintaining consistent coverage for uninterrupted connectivity and enabling positioning/sensing tasks. While achieving uniform mmWave downlink coverage in 2D scenario remains challenging yet feasible through AP wide-beam designs, 3D environments introduce significantly heightened complexity. As shown in Fig.1 (b), by deploying a planar OpenRIS on the floor sign with fine tune profile, the incident energy from narrow beam of the AP can be effectively redistributed to uniformly "illuminate" both ascending and descending staircase.

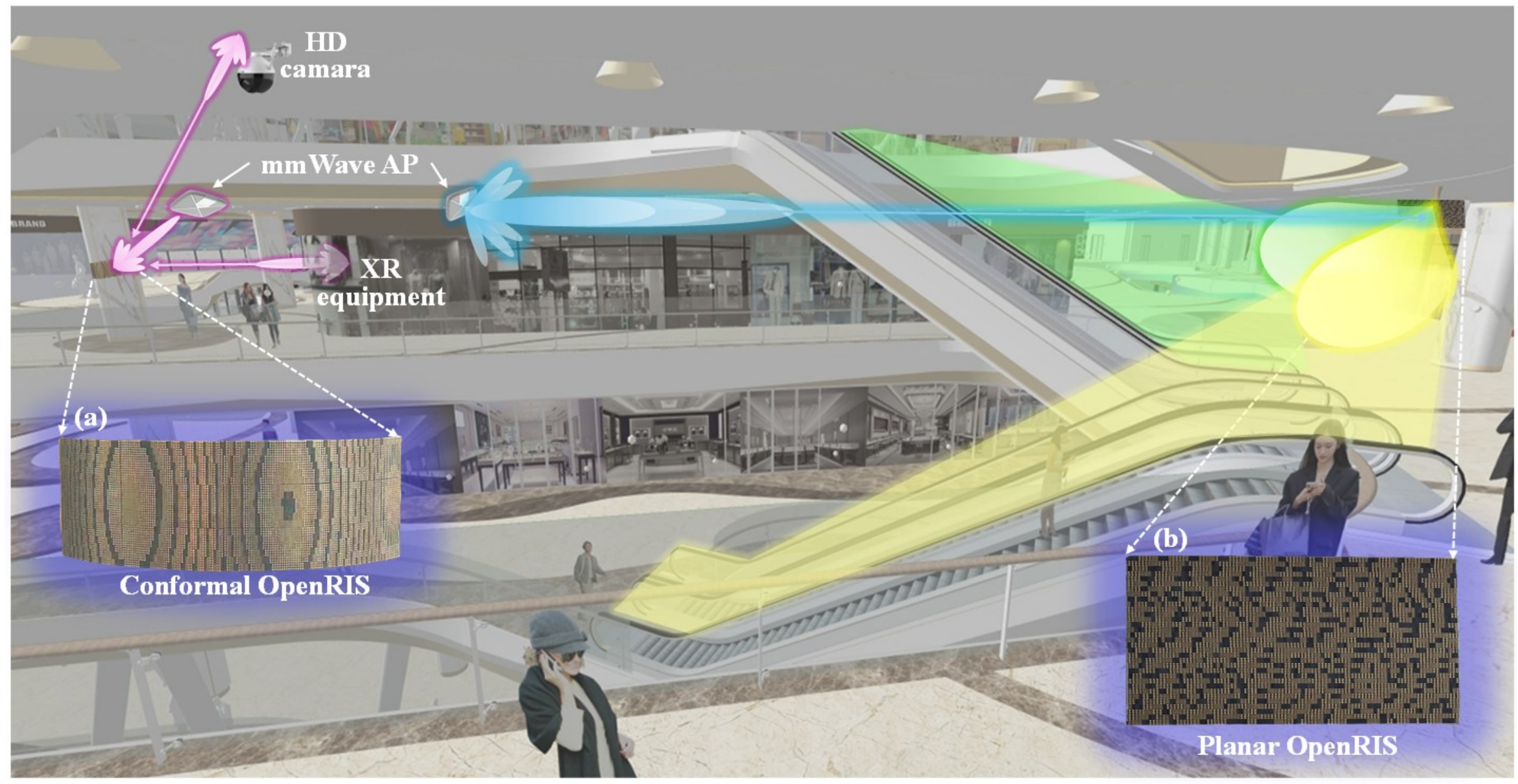

**Fig.1 Conceptual illustration of OpenRIS enabled wireless enhancement.** (a) Conformal OpenRIS steers beams for directional hotspot uplink coverage. (b) Planar OpenRIS scatters the narrow AP beam into a diffuse field, uniformly covering the 3D space traversed by ascending and descending stairs.

While the theoretical frameworks for analyzing coverage problem and designing corresponding wireless enhancement schemes are well-established, a critical gap persists in translating these methods into experimental validation across diverse scenarios due to the lack of adaptable hardware platforms. Beyond above scenario-specific coverage limitations, common constraints emerge across different environment: power supply limitations and cost sensitivity preclude the use of large-scale active RISs [37]-[40], while the architectural adaptability requirements and the need for physically reconfigurability render purely passive solutions[41]-[44] impractical. The proposed open-source, ultra-low-cost OpenRIS enables easily accessible experimental validation of theoretical wireless enhancement schemes through its modular design, zero-energy operation, and field reprogrammable properties, democratizing practical RIS implementation for coverage optimization strategies in real-world environments.

**Design and verification results of OpenRIS**

RIS functions primarily by changing the reflection phase of incident waves. In this work, we construct a 2-bits OpenRIS at 28 GHz for real-world coverage enhancement. The design and fabrication files are provided in Supplementary files. Constituted by four kinds of distinct elements, OpenRIS achieves wireless coverage enhancement via physical element assembly. As shown in Fig 2(a), the element is composed of two 0.035 mm thick copper layers separated by a F4B dielectric substrate layer with dielectric constant and loss tangent being 2.65 and 0.001, respectively. The bottom layer of element is a full-copper ground plane, while the top layer is patterned with distinct geometries characterized by $r$ and $d$ to impart tailored electromagnetic reflections. As 28 GHz, the period length and thickness of elements are 4 and 1 mm, respectively. To realize 2-bit reconfigurable reflection phase, we set $r = \{0.1, 1.2, 1.4, 1.9\}$ mm and $d = \{0.1, 0.5, 0.3, 0.2\}$ mm to obtain four elements, denoted by '0', '1', '2', and '3', whose reflection phases are approximately 0°, -90°, -180°, and -270°, respectively. Fig. 2(b) shows the full-wave simulations of these four elements using commercial software CST Microwave Studio. The designed four elements exhibit reflection amplitude losses below -0.1 dB, demonstrating excellent reflection efficiency. At 28 GHz, their phase differences are approximately 90°, enabling 2-bit phase shift. Notably, within the 27.5-28.5 GHz band, the proposed elements exhibit nearly invariant reflection loss and phase difference. Across the broader 25-31 GHz range, although some degradation in adjacent element phase difference occurs, the overall controllable phase range remains substantial. This highlights the wide-bandwidth advantage of the designed elements.

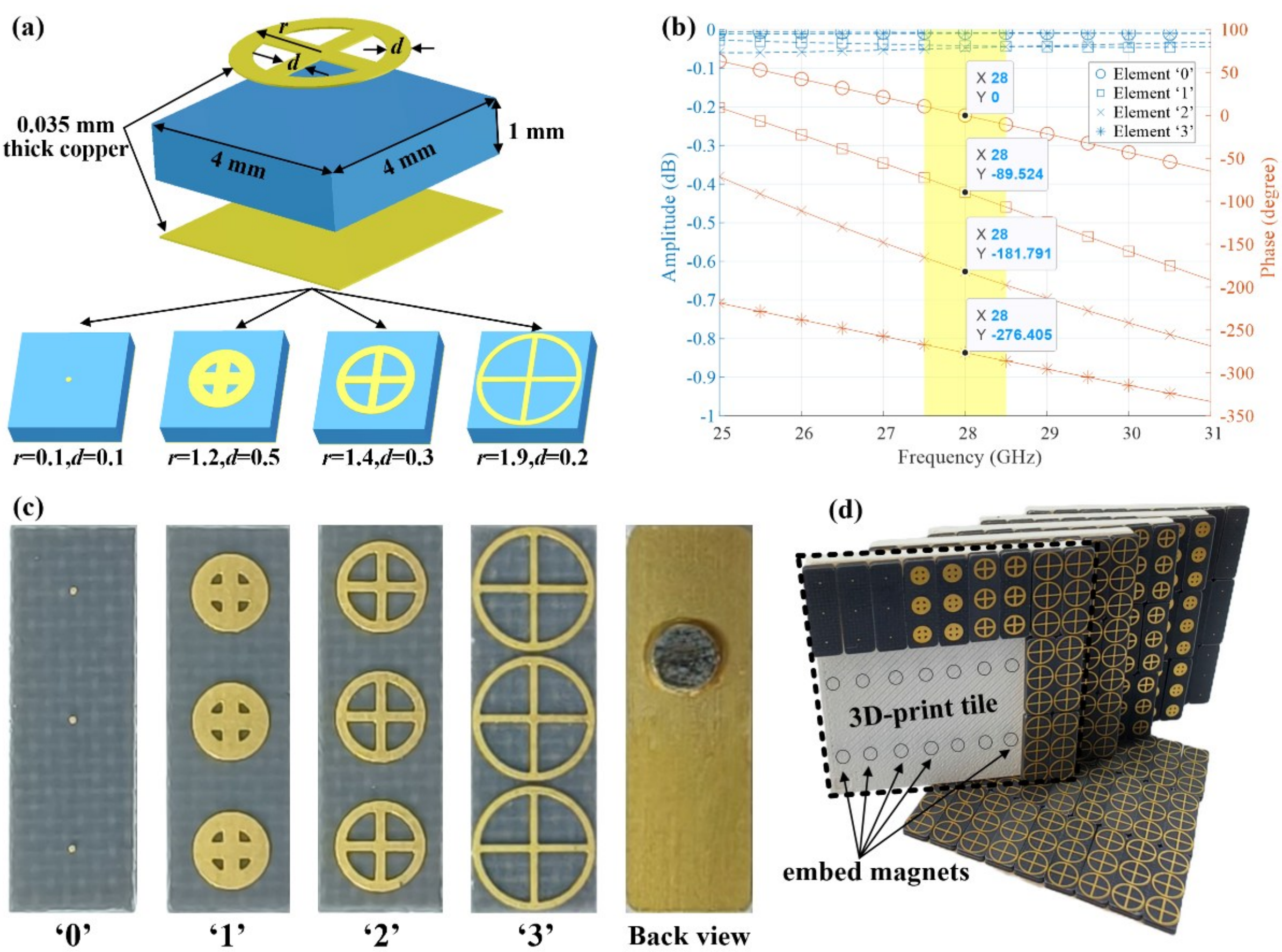

**Fig. 2 The design and fabrication of the proposed OpenRIS.** (a) Four designed elements to achieve 2-bits reflection phase shifts. (b) Simulated reflection amplitude and phase of four designed elements. (c) Photographs of the fabricated meta-bricks. Each meta-brick comprises three identical elements for manipulating the EM waves and a NdFeB magnet for snap-fit field assembly. (d) Photograph of the 3D-printed tile for rapid and accurate OpenRIS assembly with meta-bricks.

Considering the small size of mmWave element, we fabricate the meta-brick with 3×1 elements for each type, as shown in Fig. 2(c). Adopting multi-element modularization at mmWave lowers the fabrication complexity and simplifies field assembly. Furthermore, a 2 mm diameter, 1 mm thick NdFeB magnet is affixed to the bottom of each meta-bricks, allowing instantaneous snap-on/snap-off attachment to any ferromagnetic surface like building blocks. For non-metallic planes we 3D print 36 mm × 36 mm × 2 mm plastic tiles (Fig. 2(d)) that embed a regular 3 × 9 lattice of identical NdFeB magnets. This lattice precisely attracts the magnet-backed meta-bricks and eliminates manual alignment. Double-sided adhesive on other side of the tile enables rapid, large-scale deployment on arbitrary flat surfaces, making OpenRIS truly plug-and-play. Notably, the cost of OpenRIS is essentially that of PCB material itself and is less than 1% of the cost of programmable RIS[38] (see discussion in Supplementary Note 1).

As experiment validation, we assemble a OpenRIS tile that deflects a normally incident 28GHz EM wave to 42° in azimuth by arranging meta-bricks in "0-1-2-3" sequence (Fig. 3(a)). The OpenRIS tile is built from 3×9 meta-bricks snapped onto a single 3D-printed tile. Far-field measurement is performed inside a microwave anechoic chamber (Fig. 3(a)) for three prototypes: (i) 2×2 tiles (Fig. 3(b)) for a 42° wide beam, (ii) 2×4 tiles (Fig. 3(c)) for a 42° narrow beam, and (iii) 2×4 tiles (Fig. 3(d)) with reversed phase gradient for a dual 42° wide beam. Transmitter, OpenRIS, and receiver are co-polarized and co-aligned at the same height; transmitter and OpenRIS are fixed on a wooden board placed on a turntable with the OpenRIS at the rotation center. The table stepped in 0.5° increments, while the stationary receiver recorded the power radiated into each azimuthal direction, yielding the measured radiation patterns. Fig. 3(e)-(g) compare full-wave CST simulations with measurements at 28GHz (the measured results of these prototypes within the 27.5-28.5 GHz band are provided in Supplementary Note 2). For the 2×2 tile array, Fig. 3(e) shows excellent beam-steering: simulated and measured main lobes coincide almost perfectly. This result also confirms that the discrete meta-bricks lattice reproduces the behavior of a continuous aperture, thus validating the OpenRIS concept. Repeating the 2×2 tile array along the deflection direction with the same phase gradient produces a narrower beam depicted in Fig. 3(c) and (f), whereas an opposite gradient generates the dual-beam pattern in Fig. 3(d) and (g). Measurement results agree with full-wave simulations across virtually the entire angular range, with some discrepancy near 0°—where the reassured level exceeds the simulated one—being attributable to residual back-lobe radiation of the transmit horn antenna.

**Conformal OpenRIS steers beams for uplink directional hotspot coverage**

For planar RIS arrays, the most common functionality—beam steering—is employed in the preceding section to validate the effectiveness of meta-bricks. In practical deployment scenario, however, RIS implementations are not limited to planar surfaces. The proposed OpenRIS offers on-site reconfigurability to achieve specific functionalities, enabling rapid conformal deployment on arbitrary curved surfaces through integration with digital twin technology.

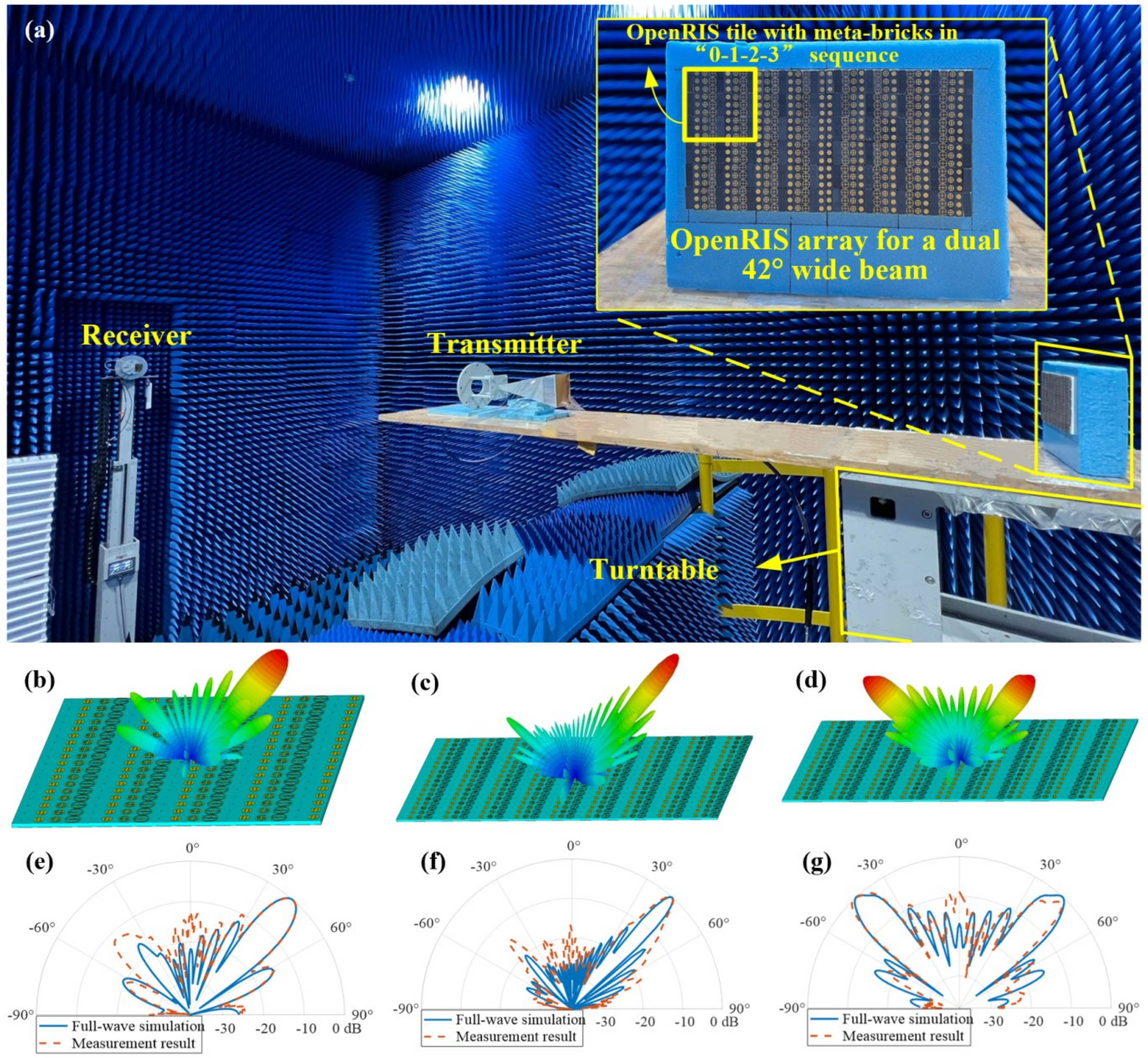

**Fig. 3 Experimental verification of the OpenRIS in microwave anechoic chamber.** (a) Photograph of the experimental setup, where transmitter, OpenRIS, and receiver are placed at the same height; transmitter and OpenRIS are fixed on a wooden board which is rotated in 0.5° steps about the RIS center. (b)-(d) Simulated reflection beam of three OpenRIS array prototypes at 28GHz: 2×2 tiles for a 42° wide beam, 2×4 tiles for a 42° narrow beam, and 2×4 tiles for a dual 42° wide beam. (e)-(f) Comparison of simulated and measured results of three OpenRIS prototypes at 28GHz.

In a scenario illustrated in Fig. 4(a), two hotspot UEs requires HD video upload to an AP. The mmWave AP, constrained by limited coverage, fails to establish high-speed links with UEs located at large angular directions. Thus, we try to assemble a conformal OpenRIS on a cylindrical pillar nearby to redirect signals from both UEs toward the AP. To realize this goal, a digital twin of this communication scenario (Fig. 4(b)) is constructed using the 3D Scanner App on an iPhone and the Sionna Ray Tracing, streamlining OpenRIS deployment on non-planar surface. Specifically, a 54×184-element conformal OpenRIS array fully adhering the pillar geometry is constructed with the digital twin, and further assembled and deployed on real world. This user-friendly and low-barrier deployment process—from the real-world, to a virtual digital twin, and back to the real-world—is demonstrated in Supplementary Video 1. Figs. 4(c) and (d) demonstrate the AP's wireless coverage before and after OpenRIS deployment in the digital twin, highlighting significant directional coverage enhancement.

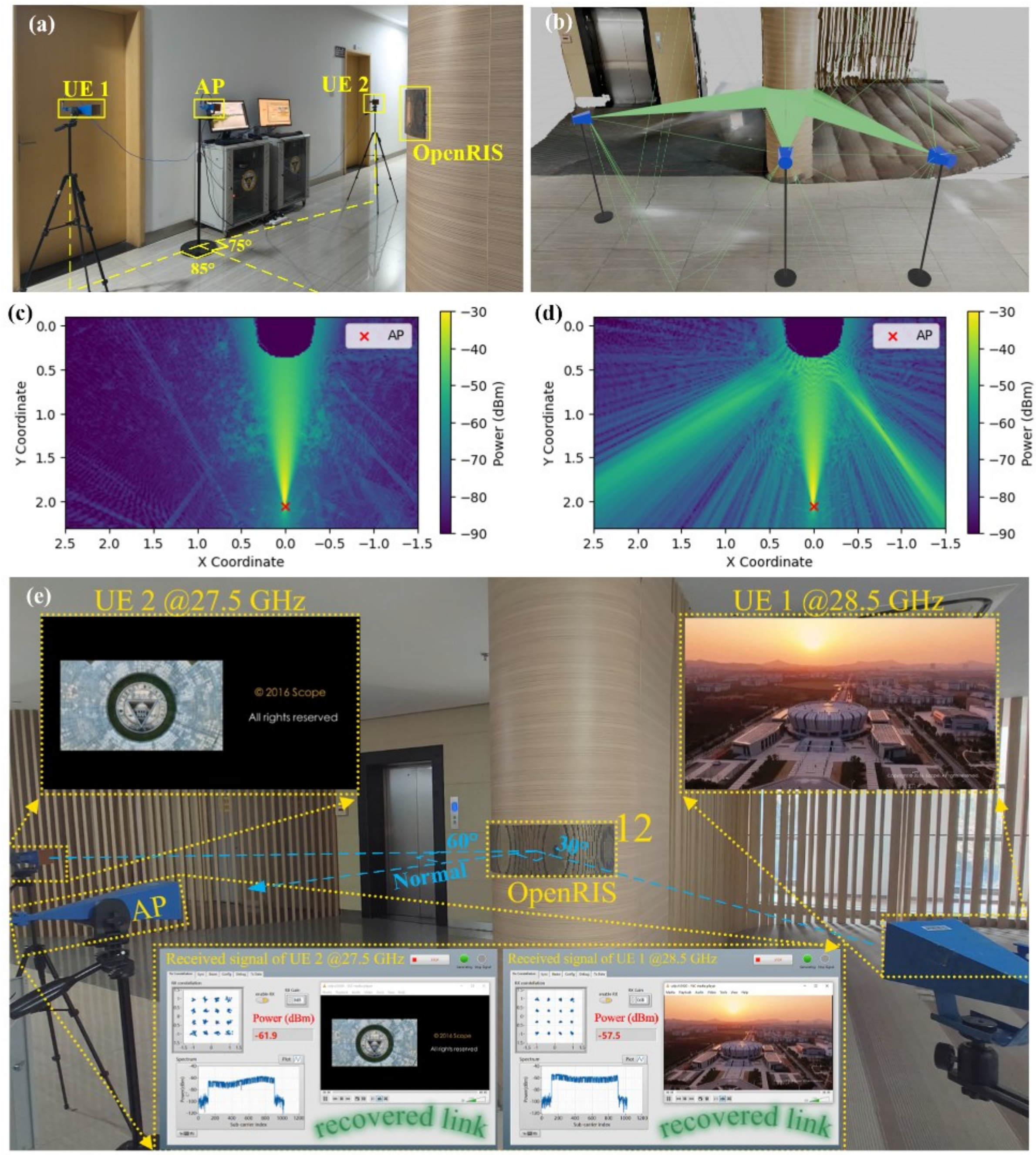

**Fig. 4 Experimental of directional uplink hotspot enhancement with a conformal OpenRIS.** (a) Photograph of the uplink hotspot transmission scenarios, where a conformal OpenRIS is deployed on a pillar to establish stable uplinks for two UEs outside the coverage of the AP. (b) A digital twin of the considered scenarios, constructed by 3D Scanner APP on iPhone and Sionna Ray Tracing. (c) Initial wireless coverage quality of the AP in digital twin. (d) Directional hotpot coverage enhancement with the OpenRIS deployment in digital twin. (e) Real-world measurements in a two-UE uplink wireless system with the assistance of the conformal OpenRIS. The uplink video transmission links are successfully established, where an approximate 20 dB improvement in received signal power is achieved for both UEs.

To experimentally validate the conformal OpenRIS's uplink performance, a two-UE wireless video transmission system is implemented (Fig. 4(e)): the AP is positioned 1.7 m along the OpenRIS normal direction, while UE 1 and UE 2 are placed 2 m from the OpenRIS at 30° and 60° angular offsets, respectively. To showcase OpenRIS's broadband capability, UE 1 and UE 2 transmit video streams centered at 28.5 GHz and 27.5 GHz, respectively, with the AP demodulating signals in corresponding bands. Prior to RIS deployment, UE signals reach the AP mainly via irregular scattering from the pillar, resulting in severe path loss and insufficient received power for video demodulation. We incrementally attached the OpenRIS tiles to positions on the pillar, which are predetermined in the digital twin, progressively improving received signal power at the AP (see Supplementary Video 2). Upon full deployment of 12 tiles, the AP achieves received signal of -57.2 dBm (UE 1) and -61.7 dBm (UE 2), corresponding to about 20 dB gains over the scenario without the OpenRIS (see Supplementary Note 3), successfully enabling stable video uploading. Notably, OpenRIS operates without active signal emission and requires no additional energy input or maintenance, merely redirecting otherwise dissipated electromagnetic waves to sustain wireless links. This deploy-and-forget operation mechanism represents a sustainable low-carbon approach that can reduces the power consumption of wireless communication systems. For instance, with the conformal RIS assistance, UE 1 can maintain stable uplink transmission even after reducing transmit power by 4 dB, matching the performance of UE 2.

**Stairwell OpenRIS turns wall into uniform scatter for downlink 3D uniform coverage**

To demonstrate the flexibility and sustainability of the OpenRIS, we reconfigure its conformal structure into a planar array to realize 3D uniform downlink coverage within stairwells. A dedicated 3D wireless coverage measurement system is further established to characterize the downlink coverage enhancement enabled by the planar OpenRIS. As illustrated in Fig. 5(a), a 99×99-element OpenRIS, composed of 11×11 planar tiles (Fig. 1(d)) snapped on a 40 cm × 40 cm iron plate, is deployed at the stairwell corner between two floors. A transmitter positioned on the upper floor normally emit narrow beam toward the OpenRIS at the same height. The elaborately assembled OpenRIS dispersed the narrow beam energy to uniformly cover ascending and descending stairs. To simulate real-world scenarios of UEs holding mobile devices, a receiver is fixed at 1.2 m height, and five test point per stair are sampled to measure received signal power via a vector network analyzer (VNA). A bare iron plate (without OpenRIS tiles) served as the control experiment to benchmark OpenRIS's performance.

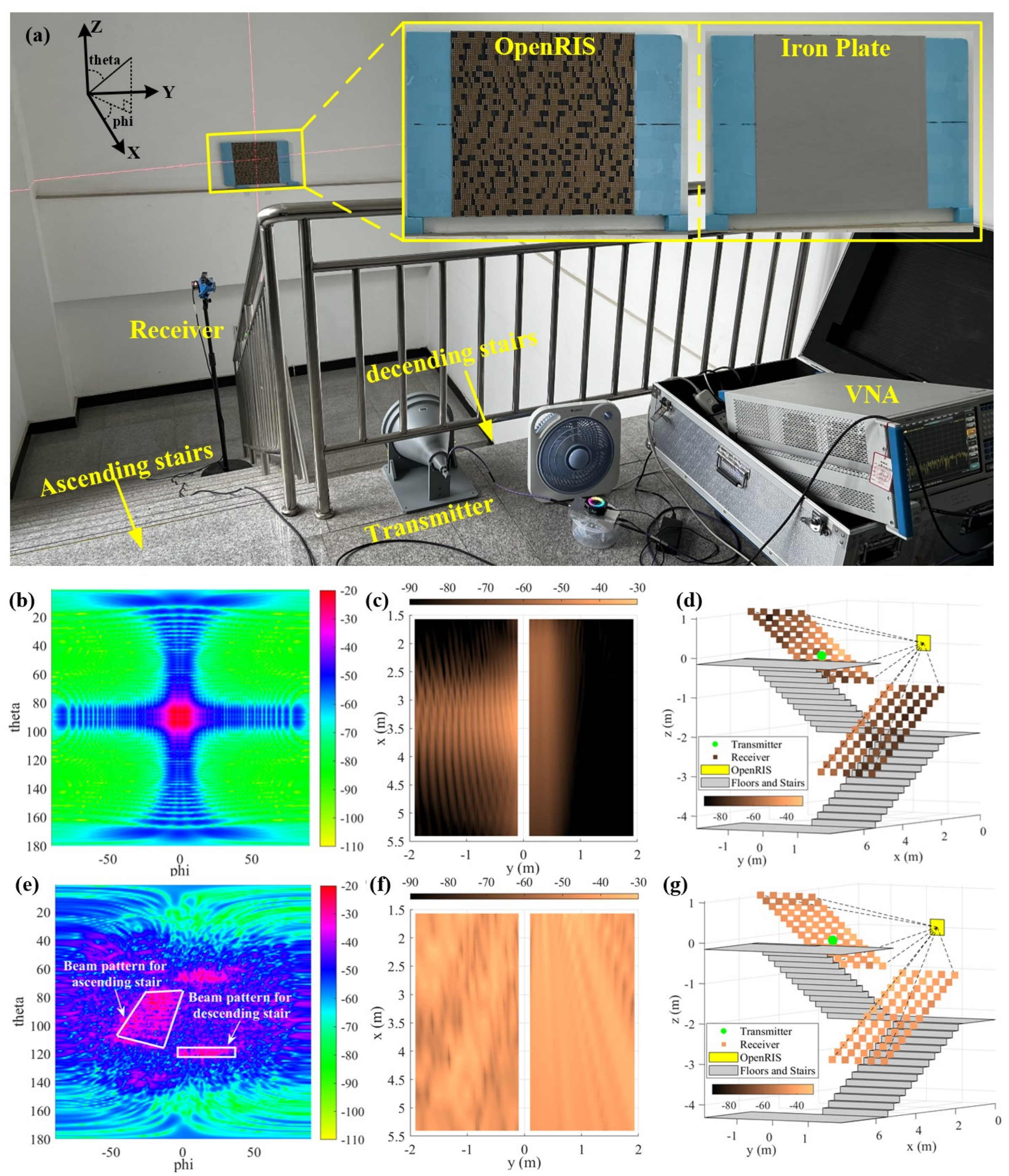

**Fig. 5 Experiments of downlink 3D coverage enhancement in stairwell with a reshaped planar OpenRIS.** (a) Photograph of the 3D coverage enhancement scenarios, where the transmitter is placed on the upper-floor aligning with the OpenRIS/iron plate deployed at the wall in stairwell corner. Five points per stair are uniformly sampled to measure the received signal via a horn antenna connecting to the VNA. (b) Reflected beam pattern of the iron plate. (c) Free-space simulation of the stairwell coverage with the iron plate. The averaged received signal power is -78.7 dBm. (d) Real-world measurements of the stairwell coverage with the iron plate. The averaged received signal power is -61.3 dBm. (e) Reflected beam pattern of the OpenRIS. (f) Free-space simulation of the stairwell coverage with the OpenRIS. The averaged received signal power is -40.4 dBm. (g) Real-world measurements of the stairwell coverage with the OpenRIS. The averaged received signal power is -41.9 dBm, an approximate 20 dB improvement compared with the iron plate case.

Fig. 5(b) and (e) demonstrate the simulated beam patterns of the iron plate and OpenRIS, respectively. While the iron plate reflects the incident beam specularly, the OpenRIS dispersed energy uniformly across the target coverage area. In the considered scenario, the target coverage occupies azimuth angles of -50° to 0° and elevation angles of 80° to 116° in ascending stairs, and azimuth angles of 0° to 40° and elevation angles of 118° to 122° in descending stairs. Fig. 5(e) demonstrate that the OpenRIS provides substantial beamforming gain within these regions. For ascending stairs, higher test points experience greater pathloss due to the increased distance; thus, the beam pattern is designed to increase gain progressively with elevation. For descending stairs, a narrow beam at 120° elevation is generated, though vertical control limitations introduce a mirrored beam due to the three-element meta-brick fabrication.

Free-space simulations of 50 uniformly sampled points per stair (Fig. 5(c) and (f)) reveal distinct coverage characteristics. The iron plate (Fig. 5(c)) reflects strong signals only at positions aligned with the specular reflection paths, while OpenRIS (Fig. 5(f)) enhance received signal strength across the coverage area, achieving an average of -40.3 dBm. Three-dimension measurements (Fig. (d) and (g)) confirm experimental consistency with simulations: the OpenRIS delivers -41.9 dBm (closely matching the simulated -40.4 dBm), while the iron plate yields a mean received power of -61.3 dBm (vs. -78.7 dBm in free-space simulation with discrepancies attributed to additional wall reflections and random scattering, which is discussed in Supplementary Note 4). The 20 dBm improvement between Fig. (d) and (g) underscores OpenRIS's efficacy in enhancing stairwell downlink coverage.

## Discussion

Improving of wireless communication performance faces greater challenges than wire systems due to the inherent reliance of open-space propagation through uncontrollable environment. The efficacy of OpenRIS in enhancing wireless coverage fundamentally depends on its phase shift capability, reflection amplitude, operational frequency bands, and transmission types. To further improve OpenRIS's performance and applicability, several potential advancements are proposed: (a) more precise phase control. The beam manipulating capability RIS is determined by the number of phase quantization bits. In this work, four element designs achieve 2-bit phase shifts. Future iterations could incorporate diversified element patterns to enable higher-bit phase control. Notably, increasing quantization bits in our proposed OpenRIS only requires fabrication additional element types, circumventing the cost escalation associated with PIN diode-based RIS or ADC-intensive varactor-based RIS; (b) more compact integration. To facilitate experimental assembly and magnetic attachment, three identical elements are group into a minimal brick for mmWave operation. This constraint introduces an upward mirrored beam during stairwell deployment. For scenarios demanding precise beam steering, adopting single-element fabrication would eliminate parasitic reflections; (c) amplified reflection. RIS-assisted systems inherently suffer from multiplicated path loss. While our fully passive OpenRIS addresses this issue through large-scale arrays, size-constrained outdoor deployment could integrate solar-harvesting modules and reflection amplifiers to boost gain without enlarging the size of OpenRIS; (d) multi-band prototypes. Given the diversity of wireless frequency bands, more OpenRIS samples optimized for distinct regimes can be designed to boarded its applicability across heterogeneous networks. (e) multimodal functionality. While the presented OpenRIS is reflective-type, transmissive or hybrid trans-reflective type designs could better serve outdoor-to-indoor penetration or multi-room coverage.

In summary, we have proposed an open-source, low-cost, and easy-to-deploy platform to democratize RIS technology in real-world wireless enhancements. The replaceable Lego-like modular architecture and zero-energy operation distinguish OpenRIS from passive fixed metasurfaces and active programmable counterpart. Experimental validation include three key phases: (i) far-field pattern measurements of small-array samples in an anechoic chamber, confirming the functionality of the designed meta-bricks; (ii) digital-twin-assisted fast deployment in a curved-wall corridor for directional hotspots enhancement, successfully establishing reliable uplinks for two UEs that is out of the coverage of an AP; (iii) stairwell deployment to uniformly distribute the narrow-beam energy into 3D coverage space, achieving an average received signal power gain of 20 dB at only 1% of the cost of conventional RIS. Importantly, OpenRIS operates as a deploy-and-forget solution for wireless enhancements, requiring no power supply or maintenance while recycling otherwise dissipated electromagnetic energy. This approach not only accelerates real-world RIS applications but also aligns with green communication paradigms by minimizing operational carbon footprints.

## Methods

### Configuration design for Conformal OpenRIS

In the conformal deployment scenario, where OpenRIS is mounted on a cylindrical pillar, the coordinate origin is set as the center of the cylinder, 1.2 meter above the ground. In the constructed digital twin, we estimate the pillar radius to be approximately 355 mm. Based on this estimation, we construct the conformal OpenRIS with a radius of 355 mm, arc length of $2\pi/3$, and height of 72 mm. The conformal OpenRIS array comprises 3×4 arc-shaped tiles, each incorporates 6×46 meticulously arranged meta-bricks (see Fig. 6(a)). Within the digital twin framework, the coordinates of the OpenRIS element, UE 1, UE 2, and AP are denoted by $\mathbf{p}_{mn}$ (element in the $m$-th row and $n$-th column), $\mathbf{p}_1$, $\mathbf{p}_2$, and $\mathbf{p}_\mathrm{a}$, respectively. Since two UEs are located on different sides of the line between the OpenRIS and AP, we divide the OpenRIS into two halves, each steering the corresponding UE's signal toward the AP. Specifically, the first $N/2$ columns of elements are assigned to UE 2 and the last $N/2$ columns for UE 1. Under this partition, we independently derive the continuous reflection phases that maximize the received-signal power of each UE and then quantize them to the 2-bit discrete set supported by meta-bricks. The procedure is detailed in the Supplementary Note 5.

### Configuration design for stairwell OpenRIS

In the stairwell scenario, the OpenRIS is deployed with its mounting wall aligned to the YOZ plane and the deploy center positioned at the origin of the Cartesian coordinate system. Within this framework, the coordinates of the OpenRIS array and its individual elements are defined as $\mathbf{p}_0 = [0,0,0]$ and $\mathbf{p}_{mn}$ (for the element in the $m$ -the row and $n$ -th column), respectively. The transmitter, placed at x-axis 5.26 meters from the OpenRIS, is denoted by $\mathbf{p}_\mathrm{t}$. To optimize the configuration of OpenRIS for uniform coverage across staircases, 50 receiver points are uniformly sampled per stair step in simulations. With 13 steps per staircase (ascending/descending), this sampling yields 1300 total receiver positions denoted by $\mathbf{p}_i$ ($i \in \{1,2,\ldots,1300\}$). The optimization objective aims to redistribute the incident narrow-beam energy from the transmitter uniformly across all ascending and descending stairs. For each $\mathbf{p}_i$, an optimal reflection phase matrix maximizing received signal power is first derived. After that, a reflection phase that enable the OpenRIS to cover all 1300 sampling points is obtained by summing the weighted reflection coefficient and extracting the phase components. To achieve uniform coverage, we widen the reflection beam[45][46] and apply a genetic algorithm to optimize the weights with the cost function being the variance of received power across all points. Algorithmic implementation details are provided in the Supplementary Note 6.

**Details on wireless video transmission system**

As shown in Fig. 6(a), the conformal OpenRIS-assisted 2-UE mmWave communication system primarily comprises three components: mmWave transceivers (UE and AP), a White Rabbit (WR) clock synchronization module, and the conformal OpenRIS array. The mmWave transmitters, UE 1 and 2, utilize the software-defined radio (SDR) platform (NI USRP-2974) to process the source video into bitstreams, which undergoes quadrature amplitude modulation (QAM) and orthogonal frequency division multiplexing (OFDM) modulation (Supplementary Note 7) to produce the digital baseband signal. This signal is converted to the analog domain, up-converted to an intermediate frequency (IF), and subsequently transmitted via a horn antenna after further up-conversion and power amplification by a mmWave frequency converter. Without direct links, the emitted signal reaches the mmWave receiver (AP) via the reflection of the conformal OpenRIS. The mmWave receiver captures the radio frequency (RF) signal through a horn antenna, where it is first subjected to low-noise amplification and down-conversion by the mmWave converter. The resulting signal is processed by the AP -side USRP-2974 for baseband conversion, channel estimation, channel equalization, and symbol detection to recover the raw video bitstreams. Subsequently, the data are decoded for video playback, and the link quality is evaluated via received signal power and constellation diagrams. To enable flexible operation across different mmWave frequencies, the local oscillator (LO) frequency of the converters is fixed at 25 GHz, with the carrier frequency adjusted via the IF signal (2.5 to 3.5 GHz). Furthermore, the synchronization module consists of a WR Switch and WR Len nodes connected via optical fiber, which distributes a precise 10 MHz reference signal to both the USRPs and mmWave converters to mitigate carrier frequency offsets and sampling time offsets.

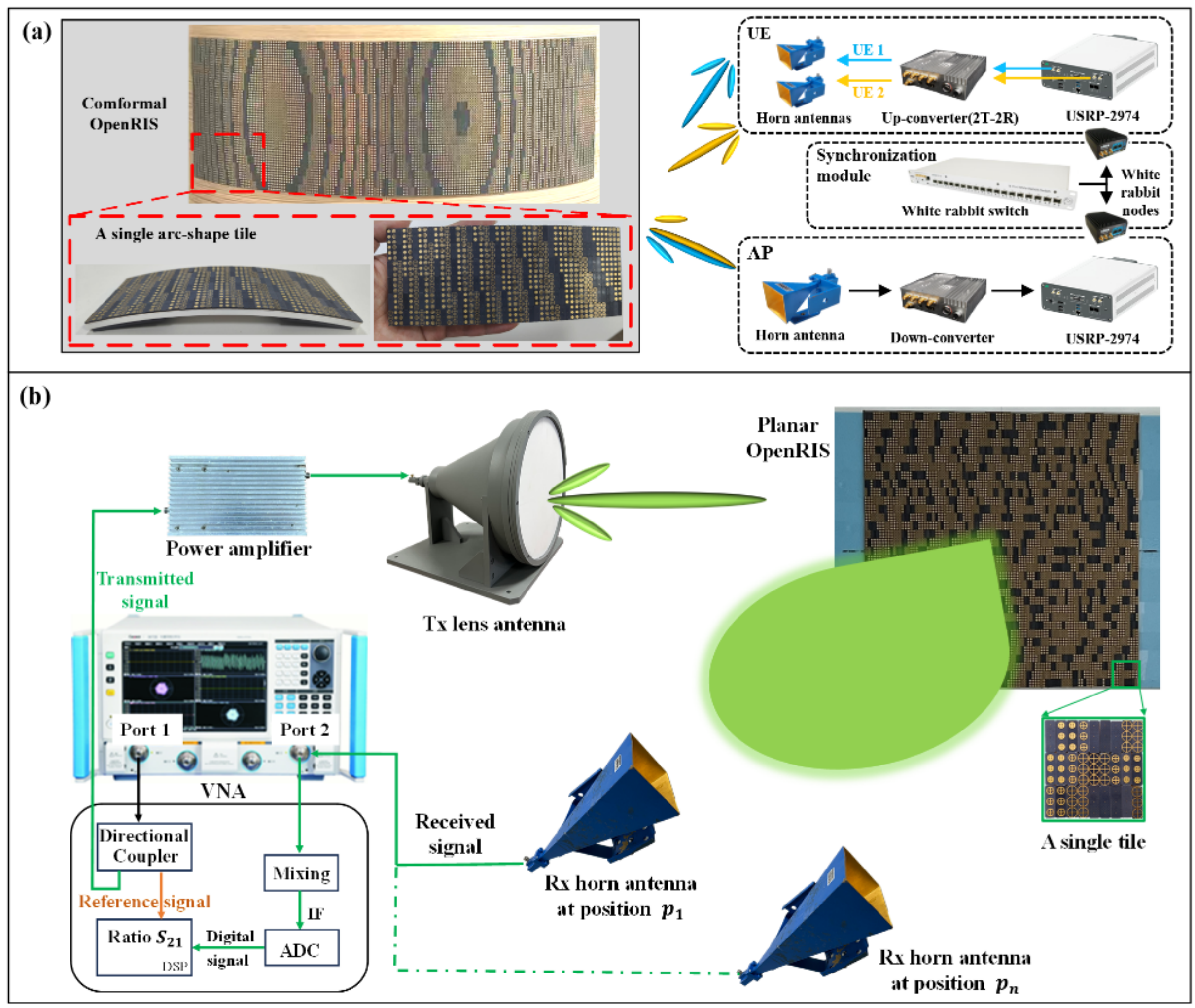

**Fig. 6 Experimental implementation of the OpenRIS-assisted systems.** (a) The conformal OprnRIS-assisted 2-UE mmWave uplink communication systems. The mmWave transceiver comprises two USRP-2974 RF chains configured as UE transmitters and one AP receiver. A WR module distributes a precise 10 MHz reference clock and pulse-per-second (PPS) signals via optical fibers to synchronize the distributed transceivers. The conformal OpenRIS redirects the signals from UEs to the AP. The fabricated OpenRIS array is constituted by 3×4 arc-shaped tiles. Each single tile, with a radius of 355 mm, arc length of π/6, and height of 72 mm, incorporates 6×46 meticulously arranged meta-bricks. (b) The planar OpenRIS-assisted coverage enhancement system. The system comprises a transceiver link consisting of a VNA, lens antenna, horn antenna, and power amplifier. By analyzing the forward transmission parameters $S_{21}$ of the receive signal relative to the reference signal, the VNA evaluates the signal quality both before and after the deployment of the planar OpenRIS.

**Details on 3D wireless coverage measurement system**

The proposed 3D wireless coverage measurement system for the stairwell primarily comprises a mmWave transceiver module and a planar OpenRIS, as shown in Fig. 6(b). The transceiver module includes a vector network analyzer (VNA) platform (Ceyear 3672E), a power amplifier, a single convex lens antenna, and a horn antenna. In operation, the VNA internally generates a 28GHz sinusoidal excitation signal, which is separated into reference and transmit measurement components by a directional coupler. The measurement signal travels from Port 1 through a power amplifier and is subsequently radiated into the environment via the lens antenna. This antenna, featuring a gain of 35.5 dBi and a 3dB beamwidth of 3°, is fixedly mounted at an elevated position facing the OpenRIS. On the receiver side, after scattered by the OpenRIS, the signal is captured by a horn antenna with a gain of 24.5 dBi and a 3dB beamwidth of $10^{\circ}$ at various sampling points and returned to VNA Port 2. Within the VNA, the receive signal undergoes down-conversion to an IF via heterodyne mixing to strictly preserve amplitude and phase integrity. Subsequently, the IF signal is digitized by an analog-to-digital converter (ADC). Finally, the digital signal processor (DSP) calculates the complex ratio of the digital measurement signal to the reference signal, thereby obtaining the forward transmission parameter $S_{21}$ for received signal power calculation.

## Data availability

The data that support the results of this study are available within the manuscript and its Supplementary Information. The design and fabrication file of the OpenRIS are provided in Supplementary files.

## Author contributions

W. C., X. L., W. X. J., T. J. C., and S. J. conceived the research idea. W. C. designed the scheme and performed the simulations and experiments in coordination with L. B., J. A., W. J. T., and W. T.. W. C. and L. B. carried out the theoretical analysis and developed the meta-bricks. J. A., X. T., W. C., and W. T. designed and built the wireless video transmission system and the 3D coverage measurement system. W. C. and W. T. conducted the system measurements with the help of J. A., X. T., and W. J. T. W. C. and L. B. collected and analyzed the data with contributions from all co-authors. W. C., W. X. J., T. J. C., and S. J. co-wrote the manuscript with input and comments from all the authors. X. L., T. J. C., and S. J. supervised and coordinated the project.

## Competing interests

The authors declare no competing interests.

**Correspondence** and requests for materials should be addressed to Xiao Li or Shi Jin or Tie Jun Cui